\documentclass[
    10pt,
    aps,
    prl,
    twocolumn,
    groupedaddress,superscriptaddress,
    footinbib,
    floatfix,
    longbibliography
    ]{revtex4-2}

\usepackage{amssymb,amsmath,graphicx,color}
\usepackage{bm}
\usepackage[tight]{subfigure}
\usepackage[export]{adjustbox}
\usepackage[bookmarksnumbered,            
            colorlinks = true,
            linkcolor = black,
            urlcolor  = blue!60!cyan,
            citecolor = blue,
            anchorcolor = blue]{hyperref}

\usepackage{lipsum}
\usepackage{multirow}
\usepackage{import} 
\usepackage[verbose]{placeins}
\usepackage{xstring}
\usepackage{bbm}
\usepackage[capitalise]{cleveref} 

\crefname{figure}{Fig.}{Figs.}
\usepackage{mathtools}
\usepackage[normalem]{ulem}

\usepackage{orcidlink}
\usepackage{comment}

\newcommand{\abs}[1]{\left\lvert #1 \right\rvert}
\newcommand{\ket}[1]{{\lvert} #1 {\rangle}}
\newcommand{\bra}[1]{{\langle} #1 {\rvert}}

\newcommand{\ccite}[1]{%
\IfSubStr{#1}{,}{refs.}{ref.~}\cite{#1}%
}
\newcommand{\Ccite}[1]{%
\IfSubStr{#1}{,}{Refs.~}{Ref.~}\cite{#1}%
}
\newcommand{\EM}{\hyperlink{endmatter}{End Matter}}

\newcommand{\beginsupplement}{%
	\setcounter{table}{0}
	\renewcommand{\thetable}{S\arabic{table}}%
	\setcounter{figure}{0}
	\renewcommand{\thefigure}{S\arabic{figure}}%
	\renewcommand{\theequation}{S\arabic{equation}}
}
\begin{document}

\title{Dissipation-assisted preparation of Floquet-Laughlin states in superconducting circuits}

\newcommand{\ITPTUB}{Institut für Physik und Astronomie, Technische Universität Berlin,
Hardenbergstr.\ 36, D-10623 Berlin, Germany}
\author{Luis C. Steinfadt}
\email{steinfadt@tu-berlin.de}
\affiliation{\ITPTUB}
\author{André Eckardt}
\affiliation{\ITPTUB}
\author{Francesco Petiziol}
\affiliation{\ITPTUB}

%%%%%%%%%%%%%%%%%%%%%%%%%%%%%%%%%%%%%%%

\begin{abstract}

Fractional Chern insulators (FCIs) are lattice analogs of fractional quantum Hall systems, where the interplay of strong interactions with a frustrated tunnelling kinetics leads to the emergence of a gapped ground state with long-range entanglement and anyonic excitations. The highly correlated nature of such systems makes their adiabatic preparation challenging already beyond the minimal system size of two particles. Considering Floquet implementations of the bosonic Harper-Hofstadter-Hubbard model of few photons in superconducting circuits, we design protocols for the driven-dissipative stabilization of its FCI ground state at half filling via quantum bath engineering. Dissipation control is achieved through the coupling to driven leaky cavity modes, which realize a tuneable artificial environment having the Floquet-FCI as its approximate fixed point. For systems of two, three and six particles, we show numerically how the flexibility of the control scheme further allows for the detection of fractional quantum Hall signatures in the stabilized steady states, including bulk incompressibility, Hall response and the trapping of fractional charges. Our results provide a concrete pathway to dissipation-assisted preparation of strongly correlated states in quantum simulators. 

\end{abstract}

\maketitle

\hypersetup{linkcolor=blue} 

Topologically ordered phases of matter are characterized by long-range entanglement~\cite{Wen2007}, which offers a pathway to encode and manipulate quantum information in an intrinsically robust way~\cite{Kitaev2003,Freedman2003,Nayak2008}. Paradigmatic examples include fractional quantum Hall systems~\cite{Laughlin1983a,Moore1991a,Read1999,DasSarma2005} and their lattice analogs, fractional Chern insulators (FCIs)~\cite{Parameswaran2013, Bergholtz2013}. In this context, the Harper-Hofstadter-Bose-Hubbard (HHBH) model at half filling has proven to be a particularly promising candidate for realizations in quantum simulators~\cite{Tai2017,leonardRealizationFractionalQuantum2023a, Rosen2024, Wang2024}. However, preparing its FCI ground state, which closely resembles a discretized version of the $\nu=1/2$ Laughlin state~\cite{hafeziFractionalQuantumHall2007}, poses an outstanding challenge. Recent experimental efforts have succeeded in detecting FCI signatures for a minimal system of two particles so far, with ultracold atoms in quantum gas microscopes~\cite{leonardRealizationFractionalQuantum2023a} and superconducting qubits~\cite{Wang2024}. This has been achieved by combining the realization of artificial magnetic flux via time-periodic driving (Floquet engineering~\cite{Eckardt2017, Bukov2015}) with an adiabatic preparation scheme. The adiabatic approach is however constrained by drawbacks that limit the attainable fidelities and challenge the applicability to larger systems~\cite{Palm2024, Blatz2024, Wu2025}: On the one hand,  it relies on the fact that, in a small system, gap closings at the topological phase transition can be circumvented through optimized multi-parameter ramps~\cite{Popp2004,He2017,Motruk2017,Hudomal2019,Andrade2021}; on the other, unavoidable excitations produced by finite-time ramping cannot be disposed of, significantly polluting target observables.

In this Letter, we propose a scheme to induce the autonomous stabilization of the FCI ground state of the HHBH Hamiltonian, based on controlled dissipation~\cite{Diehl2008,Kraus2008,Verstraete2009,Carusotto2009a,Krauter2011,Umucalilar2012,Biella2017,liuDissipativePreparationFractional2021,Carusotto2025}. Considering a Floquet HHBH lattice of superconducting qubits~\cite{Rosen2024, Wang2024}, we provide detailed recipes for how to engineer the environment through pumped leaky cavities~\cite{Murch2012,Hacohen-Gourgy2015, Kapit2017, Carusotto2020} to stabilize the FCI ground state of the effective Floquet-engineered Hamiltonian for different system sizes. Differently from previous strategies based on particle loss and refilling~\cite{kapitInducedSelfStabilizationFractional2014,Ozawa2014,Lebreuilly2016,Lebreuilly2017,Umucalilar2017,Kurilovich2022}, our approach works at conserved particle number and explicitly takes into account the interplay of Floquet and incoherent driving~\cite{petiziolCavityBasedReservoirEngineering2022, Mi2024, Schnell2024, Ritter2025}. Our results show the successful preparation of FCI states in systems significantly larger than those accessible via adiabatic state preparation, without requiring numerical optimization of parameters. Focussing on progressively larger systems of up to six particles in a $8\times 8$ lattice, we discuss how key FCI signatures in the stabilized steady states can be extracted by adapting the control protocol and show that they exhibit excellent agreement with pure-ground-state results. The examples discussed here thus provide a blueprint for the observation of signatures not yet accessible in the system sizes realized so far, such as the pinning of fractional charges.

\emph{Floquet-HHBH Hamiltonian and reservoir engineering.} Consider an $L\times L$ square array of superconducting artificial atoms described by the HHBH Hamiltonian
\begin{equation}
    \frac{\hat H_\text{\scriptsize FCI}}{J_\text{eff}} = - \sum_{m, n} e^{-i\phi n}\hat a_{m, n}^\dag \hat a_{m + 1, n} + \hat a_{m, n}^\dag \hat a_{m, n + 1} + \text{H.c.}
    \label{eq: HHBH Hamiltonian}
\end{equation}
Here, $\hat a_{m,n}$ is the hard-core bosonic annihilation operator at site $(m,n)$ and $J_\text{eff}>0$ the effective tunneling amplitude. Tunneling along the first coordinate is accompanied by a Peierls phase $n\phi$ proportional to the second coordinate. This phase effectively models a magnetic flux $\phi$ piercing the lattice, and we analyze the situation where it is Floquet-engineered by periodically modulating the atoms' transition frequency~\cite{Rosen2024} (corresponding to the on-site potentials of the Bose-Hubbard description). 
As detailed in \EM, \cref{eq: HHBH Hamiltonian} emerges in this case as the period-average of $\hat H_\text{s}(t) = - J \sum_{\langle l, l^\prime\rangle} e^{i[\theta_{l}(t)-\theta_{l^\prime}(t)]}\hat a_l^\dag \hat a_{l^\prime} + \text{H.c.}$, for on-site potentials $\hbar\dot\theta_{l}(t)=\hbar\dot\theta_{l}(t+2\pi/\omega)$ oscillating at large frequency, $\hbar\omega\gg J$~\cite{petiziolCavityBasedReservoirEngineering2022,Rosen2024}.

\begin{figure*}[ht]
    \centering
    \includegraphics[width=\textwidth]{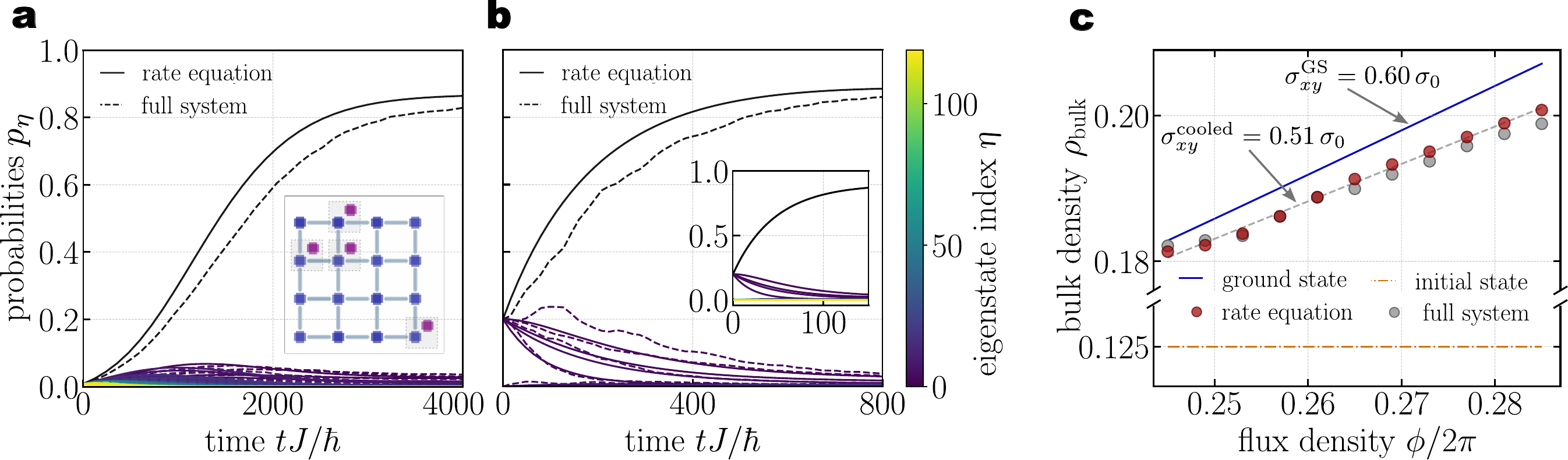}
    \caption{(a) Driven-dissipative stabilization of the ground state of the $L=4$ HHBH lattice with $N=2$ at quarter flux, $\phi=\pi/2$, starting from an infinite temperature state. A fidelity of $85\%$ is reached using four cavities. Inset: Lattice geometry with artificial atoms and cavities marked by blue and purple bullets, respectively. The solution of \cref{eq:ELE} (solid lines) and \cref{eq:PRE} (dashed) show good agreement. (b) The stabilization time can be decreased to less than $800 \, \hbar/J$, when starting from a superposition of the five lowest energy eigenstates, and to less than $200 \, \hbar/J$ (inset) by further adding twelve identical cavities, coupled to symmetrically chosen lattice sites.
    (c)  The bulk density of the stabilized mixture increases linearly with the flux $\phi$. The slope of the dashed line estimates the fractional Hall conductivity of the stabilized mixture as $\sigma^\text{cooled}_{xy}/\sigma_0 = 0.51$, close to the exact HHBH ground state value $\sigma^\text{GS}_{xy}/\sigma_0 = 0.6$. The orange dashed line gives the value of either observable for the initial (infinite temperature) state.}
    \label{fig:collage_4x4}
\end{figure*}

Dissipation engineering is implemented by coupling each of a subset of lattices sites with strength $g_j$ to a pumped, leaky resonator mode of frequency $\delta_j/\hbar$ (relative to the atoms' frequency) and decay rate $\kappa_j$, as sketched in the inset of \cref{fig:collage_4x4}a. In a nutshell, photons pumped in the resonator at a red-(blue-) detuned frequency $\omega_j=(\delta_j-d_j)/\hbar$, with amplitude $\mathcal{E}_j$, can scatter on the system and extract (dump) the missing energy $d_j$ needed for a re-emission on resonance by the resonator into the environment, resulting in cooling (heating) of the system~\cite{Murch2012, Hacohen-Gourgy2015, clerkIntroductionQuantumNoise2010,Petiziol2024b}.

For the cavities to exchange energy with the lattice while preserving its particle number, the coupling is assumed dispersive, such that the detuning between atoms and cavities is large compared to both the atom-cavity coupling and the tunneling energy in the lattice, $\delta_j \gg J, g_j$. 
As the lattice is driven strongly to Floquet engineer the HHBH Hamiltonian of \cref{eq: HHBH Hamiltonian}, the impact of the modulation on the interaction with the cavities must be taken into account~\cite{petiziolCavityBasedReservoirEngineering2022}. First, to prevent drive-assisted tunneling of particles between the lattice and the cavities, potentially induced by the modulation, the Floquet frequency $\omega$ and its integer multiples $\mu\omega$ are highly detuned from resonances activating such processes. As $J\ll \delta_j$, this approximately implies $\mu\omega\ne \delta_j/\hbar$. The modulation further leads to a dressing of the system-cavity coupling and, as a consequence, of the engineered dissipation rates, which we derive following Ref.~\cite{petiziolCavityBasedReservoirEngineering2022} -- see also \EM.

All in all, the effective dynamics of the combined system of lattice and cavities in the high-frequency and dispersive regime is given by the Lindblad master equation
\begin{equation}
\label{eq:ELE}
\hat \rho (t) = -\frac{i}{\hbar}[\hat H_\text{eff}, \hat{\rho}(t)] + \sum_j \kappa_j \mathcal{D}_{\hat c_j}[\hat \rho(t)],
\end{equation}
with dissipators $\mathcal{D}_{\hat c_j}(\cdot) =  \hat c_j (\cdot) \hat c_j^\dag - \frac{1}{2}\hat c_j^\dag \hat c_j (\cdot) - \frac{1}{2} (\cdot)\hat c_j^\dag \hat c_j$. Here, $\hat{c}_j$ denotes the bosonic annihilation operator of the cavity mode at site $j$. The effective Hamiltonian for system and cavities can be written as $\hat H_\text{eff}= \hat H_\text{FCI} + \hat H_\text{sc} + \sum_j d_j \hat c_j^\dag \hat c_j$, with atom-cavity interaction
\begin{align}
    \hat{H}_\text{sc} = - \sum_j  \chi_j \hat{a}^\dag_j \hat{a}_j \left[\hat c_j ^\dag \hat c_j + \alpha_j \hat c_j^\dag + \alpha_j^* \hat c_j \right],
    \label{eq: eff coupling}
\end{align} 
where the coupling $
    \chi_j = \sum_{\mu=-\infty}^\infty 2g^2_j \mathcal{J}^2_\mu \left( \frac{\lambda}{\hbar \omega} \right)/(\delta_j - \mu\hbar\omega)
    $
reflects the Floquet dressing through the summation over Floquet harmonics, the Bessel-function renormalization and the multi-photon resonances $\delta_j-\mu\hbar\omega$.
Equation \eqref{eq: eff coupling} is written in a frame, where the cavities are displaced to a coherent state of amplitude $\alpha_j=- \mathcal{E}_j/(d_j - i\hbar\kappa_j/2)$, which would correspond to their steady state for zero coupling to the atoms. 

To understand how the coupling to the cavities in \cref{eq:ELE} translates into controllable dissipative dynamics in the lattice, and be used to design a dissipative cascade to the FCI, we further derive the cavity-induced transition rate between eigenstates $\ket{\varepsilon_\eta}$ with (quasi)energy $\varepsilon_\eta$ of \cref{eq: HHBH Hamiltonian} (see \EM). In the `bad-cavity' regime, $\vert \varepsilon_{\eta} - \varepsilon_{\eta^\prime} \vert \gg \hbar \kappa_j \gg \vert \alpha_j\chi_j \bra{\varepsilon_\eta}\hat n_j\ket{\varepsilon_{\eta^\prime}}\vert$, the cavity relaxation is only weakly affected by the atoms and each cavity can be understood as a bath with a narrow Lorentzian spectral density centered at $d_j$ of bandwidth $\kappa_j$~\cite{clerkIntroductionQuantumNoise2010}. Tuning the pump detuning $d_j$ thus grants a handle on the energy at which each cavity stimulates emission or absorption from the atoms, with $\kappa_j$ determining the energy resolution and the imbalance between absorption and emission rates (akin to an effective temperature for the transition). After tracing out the cavities and performing a rotating-wave approximation, the lattice dynamics can then be approximated by the Pauli rate equation 
\begin{equation}
\label{eq:PRE}
    \dot{p}_\eta=\sum_{\eta^\prime}\left(R_{\eta\eta^\prime}p_{\eta^\prime} - R_{\eta^\prime\eta}p_{\eta}\right)
\end{equation}
for the population $p_\eta$ of the many-body eigenstate $\ket{\varepsilon_\eta}$, with transitions $\ket{\varepsilon_{\eta^\prime}} \mapsto \ket{\varepsilon_\eta}$ occurring at a rate 
\begin{align}
        R_{\eta\eta^\prime} = \sum_j\frac{\kappa_j \chi_j^2 \vert \alpha_j\vert^2 \left\vert\bra{\varepsilon_\eta}\hat n_j\ket{\varepsilon_{\eta^\prime}}\right\vert^2}{(\varepsilon_\eta-\varepsilon_{\eta^\prime} + d_j)^2 + \hbar^2\kappa_j^2/4}.
        \label{eq: pauli rates}
\end{align}
We show in the following how, guided by Eq.~\eqref{eq: pauli rates}, the cavity configuration and the control parameters can be chosen to create a dissipative path that autonomously drives the system to the FCI from generic initial states, for systems of two, three and six particles.

The numerical results presented below are obtained, for the two-particle FCI, by solving \cref{eq:ELE} for the combined lattice-cavity system though quantum trajectories~\cite{Weinberg2019,Lambert2026}. To address larger systems, we solved \cref{eq:PRE} for the eigenstate populations in a low-energy subspace, whose basis states are computed via either exact diagonalization, for three particles, or matrix-product-state ans\"atze and excited-state DMRG~\cite{Hauschild2018,Catarina2023}, for six particles. The parameter values used are discussed in the \EM\ and reported explicitly, for each figure, in the Supplementary Material~\footnote{See Supplementary Material, where tables reporting the parameters values used in the simulations producing all figures can be found.}.

\emph{Two-particle FCI}. We first analyze the $L=4$ lattice with $N=2$ particles, recently realized experimentally~\cite{leonardRealizationFractionalQuantum2023a,Wang2024}. Assuming a worst-case scenario, the lattice is initialized in an infinite-temperature state. Using four cavities, coupled to four different sites as shown in the inset of \cref{fig:collage_4x4}a, the HHBH ground state is stabilized to a fidelity larger than 85\% in an evolution time of $4000 \, \hbar/J$ (\cref{fig:collage_4x4}a). We shall discuss below, how this timescale can be reduced drastically by using more cavities and leveraging lattice symmetries. These results were obtained by setting the detunings of the first three cavities to $d_j/J \approx 0.125, 0.161, 0.273$, namely, resonant with the energy differences between the ground state and the first three excited states, while the last cavity addresses larger transitions ($d_j \approx 0.4J$). The chosen cavity arrangement also ensures large coupling matrix elements $\bra{\varepsilon_\eta} \hat n_j \ket{\varepsilon_{\eta^\prime}}$.

The cavity decay rate $\kappa_j$, the cavity-pump amplitude $\mathcal{E}_j$, cavity-atom coupling $g_j$ and detuning $\delta_j$ were selected to comply with ranges reported in circuit QED experiments~\cite{Murch2012, Hacohen-Gourgy2015, Rosen2024, Wang2024, Blais2021} and ensuring that neither the assumptions of the Floquet and reservoir engineering nor of the bad-cavity regime were violated. In those regimes, the cavity arrangement and parameter values were set empirically to maximize the rates between low excited states and the ground state, given by \cref{eq: pauli rates}. Generally, this already results in population almost fully cascading into the ground state. Minor additional fine-tuning is then used to enhance the decay rates of excited states accumulating the remaining occupation, further improving fidelities. 

The successful stabilization starting from the equal-weight mixture underlines the robustness of the reservoir engineering scheme. In practice, stabilization times can be decreased drastically if the system is already initialized in a low-energy state. This may be the case, for instance, if the engineered cooling is activated after an imperfect adiabatic preparation scheme, as the ones used in the experiments of~\Ccite{leonardRealizationFractionalQuantum2023a,Wang2024}. For an equal-weight superposition of the five lowest energy eigenstates, we achieve stabilization to the same fidelity in only $800 \,\hbar/J$ (cf. \cref{fig:collage_4x4}b). This time can be further reduced to $200 \,\hbar/J$ by exploiting the fourfold rotational symmetry of the HHBH lattice and adding copies of the four cavities in a symmetric fashion (\cref{fig:collage_4x4}b--inset).

While the stabilized mixture features a large ground-state fidelity, it also includes a minor contamination by low-energy excited states. We demonstrate next that this does not noticeably pollute FCI observables, confirming that the proposed scheme provides a robust gateway to FCI physics. This also showcases the robustness of our driven-dissipative scheme in the preparation and control of correlated states. 
To this end, we probe the fractional Hall conductivity $\sigma_{xy}$ in the two-particle FCI. It can be obtained from the response of the bulk density $\rho_\text{bulk}$ of the stabilized mixtures to a variation in the magnetic flux $\phi$, as predicted by St\v{r}eda's formula, $\sigma_{xy} = \partial_\phi \rho_\text{bulk}$~\cite{Streda1982,repellinFractionalChernInsulators2020}. We find a steady-state fractional Hall conductivity $\sigma^\text{cooled}_{xy}/\sigma_0 \approx 0.51$, closely matching the value expected in the thermodynamic limit, $\sigma^\text{TDL}_{xy}/\sigma_0 = 0.5$, and the exact value $\sigma^\text{GS}_{xy}/\sigma_0 \approx 0.6$ found for the ground state of the finite system considered here (cf. \cref{fig:collage_4x4}c). Here, $\sigma_0^{-1} = R_\text{K}$ is the von Klitzing constant. 

The above results also highlight that the dissipative preparation scheme is robust against offsets in the values of the control parameters: performances for a given protocol are barely affected by small variations of $\phi$. This robustness is confirmed also with respect to weak local on-site potential offsets $V\sim J/2$, which can be used to probe incompressibility~\cite{Raciunas2018, wangMeasurableSignaturesBosonic2022, Wang2024}, as also discussed below for larger systems. 
The small two-particle system discussed so far, for which the exact numerical simulation of \cref{eq:ELE} is computationally accessible, also allowed us to test and confirm the accuracy of the reduced Pauli rate equation~\cref{eq:PRE}, which will be used for larger systems, where \cref{eq:ELE} is too demanding.
As an example, \cref{fig:collage_4x4} shows that the transient population dynamics of \cref{eq:ELE} and \cref{eq:PRE} have good agreement and the steady state fidelities match closely. Further, the steady-state compressibility and conductivity behavior is nearly identical in the relevant regimes. Having gained trust in the rate equation, we can now employ it to study the preparation of three- and six-particle FCI states. Their realization has yet not been achieved via adiabatic preparation experimentally and grants access to a broader range of signatures of topological order.

\begin{figure}[t!]
    \centering
    \includegraphics[width=.45\textwidth]{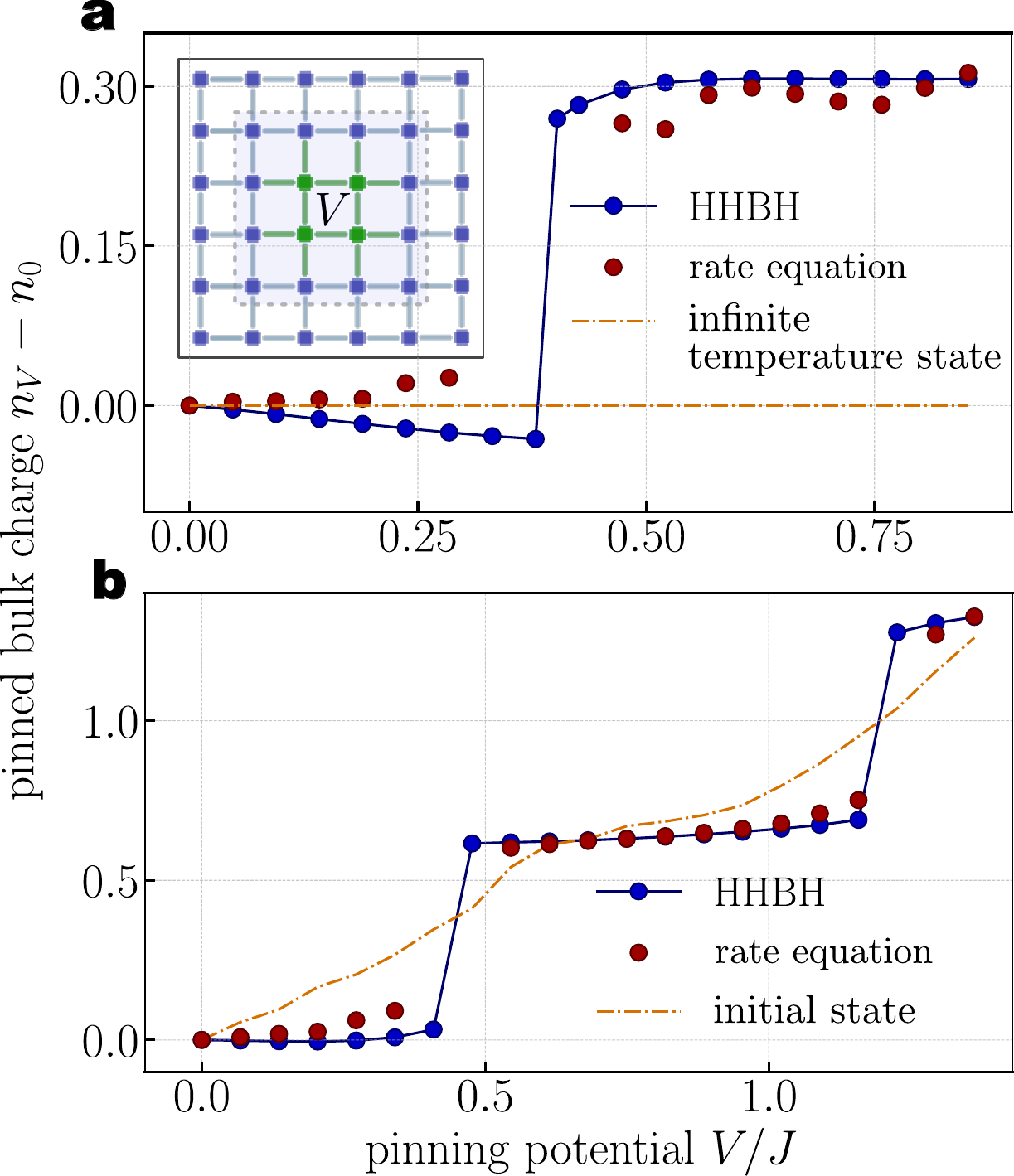}
    \caption{Change of the bulk charge, defined as the average particle number in the inner $L-1\times L-1$ sites, in response to a local potential dip of $-V$ on four sites the center of a system of $N=3$ and $L=6$ (a) or $N=6$ and $L=8$ (b). For each value of $V$, the system is initialized in a mixture of low-energy eigenstates and stabilized to the ground state, reaching a fidelity of around 80\%. The change in the total charge in the bulk compared to the unperturbed state was computed (cf. inset of (a) for the $L=6$ lattice, analog for $L=8$). In both cases, the stabilized mixture behaves incompressible for small $V$, as in the exact FCI ground state. After a critical value of $V$, the total charge in the bulk changes abruptly, hinting at the pinning of a quasiparticle in the bulk. For larger values of $V$, the system remains incompressible until another quasiparticle may be created (cf. (b) for $V\gtrsim 1.1 \,J$).}
    \label{fig:pinning}
\end{figure}

\emph{Three-particle FCI}. For the preparation of a three-particle FCI at half filling, requiring $L=6$, we study the relaxation dynamics of an equal-weight mixture of the lowest fifty energy eigenstates in the presence of a potential dip of varying magnitude $-V$ applied to the central four sites, as depicted in green color in the inset of \cref{fig:pinning}a. For all values of $V$ considered, ground-state fidelities around 75\% within $3000 \,\hbar/J$ are attained, using six cavities.
In \cref{fig:pinning}a, the variation of the total charge in the bulk in the steady state at different potential depths $V$ is shown. For $V\lesssim 0.4 J$, both the stabilized mixture and the exact ground state only respond marginally to the applied potential, reflecting the bulk incompressibility found in fractional quantum Hall states. 
For $V\gtrsim0.4J$, the potential dip overcomes the many-body gap and the ground state and first-excited state exchange character, with the latter hosting a pinned quasiparticle~\cite{wangMeasurableSignaturesBosonic2022} in the bulk. This manifests itself in both the steady state and the exact ground state remaining incompressible, but now with the particle density increased approximately by the expected quasiparticle charge of $1/2$~\cite{wangMeasurableSignaturesBosonic2022}. 
Data points for the rate equation near $V\approx 0.4\,J$, where the state with and without a trapped quasiparticle exchange character, were omitted: In this range, the engineered environment cannot resolve the two states for realistic values of $\kappa_j$ and thus stabilizes a mixture of them. Interestingly, this sensitivity of the dissipation-control scheme may be considered as a promising proxy for detecting phase competitions. 

We point out that, for three particles, stabilization times are three to four times larger than for $L=4$ in \cref{fig:collage_4x4}b, despite the relevant spectral width to be covered by the cavities being approximately the same, $\sim J/2$. 
On the one hand, this is due to the gap between ground and first excited state decreasing with the system size--- a variation attributed to finite-size effects. This, in turn, requires a smaller decay rate $\kappa_j/J \in [0.006, 0.04]$ to fully depopulate the excited state (cf. $\kappa_j/J \in [0.01, 0.1]$ for $N=2$). On the other, the density of states, and thus the number of necessary transitions to reach the ground state, increases. Both effects slow down the stabilization dynamics.

\emph{Six-particle FCI}. We finally address the stabilization of the six-particle FCI on an $L=8$ lattice. Since exact diagonalization is not possible in this case, we reconstruct a low-energy sector spanned by $d = 20$ energy eigenstates through DMRG on matrix-product states~\cite{Hauschild2018,Catarina2023} and investigate stabilization therein with \cref{eq:PRE}. Ground state fidelities $\sim 80\%$ could be achieved for different pinning potential strengths $V$. The pinning of a fractional charge in the cooled mixture agrees very well with the result obtained from the HHBH ground state, as shown in \cref{fig:pinning}b, which behaves qualitatively as in the three-particle case. 

These results provide encouraging proof of principle that our proposed cavity-assisted stabilization can successfully be applied to larger FCIs, when a relatively low-energy state is initialized. Indeed, as a caveat, it should be kept in mind that  the numerically achieved Hilbert-space truncation via DMRG may not be sufficient to fully accommodate initial states easily prepared in experiments, thus potentially leading to an over-(under-) estimation of fidelities (stabilization times). The truncation may also exclude, by construction, Floquet multi-photon resonances to higher excited states [see the denominators of $\chi_j$ in \cref{eq: pauli rates}] that compete with the cooling scheme at long times. 
Moreover, with larger system size, the driving frequency cannot be chosen anymore to be large compared to the full spectral width of the time averaged Hamiltonian $\hat H_\text{eff}$, so that resonant excitation processes, known as Floquet-heating, start to matter, which are not yet included in the description based on $\hat H_\text{eff}$. However, it was pointed out recently \cite{Wanckel2026} for the case of thermal environments that, despite the potential problems just mentioned above, gapped ground-states of $\hat H_\text{eff}$ can nevertheless be prepared with high-fidelity, provided the relevant bath-induced processes (as we capture them here) are fast compared to unwanted resonant excitation channels (given by both resonant bath-induced processes and Floquet heating). This reasoning applies also to the engineered reservoirs considered here, making them in principle scalable to even larger system sizes. 

\emph{Conclusion}. We showed how the coupling of a Floquet-engineered Harper-Hofstadter-Bose-Hubbard lattice to driven and leaky cavities can be engineered and used to stabilize small-scale fractional Chern insulator states of photons in a superconducting circuit, for system sizes ranging from two to six hardcore particles. Our driven-dissipative approach provides a robust means to both prepare such correlated states and probe their properties. Indeed, we tested schemes to probe FCI signatures in the stabilized steady states, confirming clear evidence of bulk incompressibility, fractional Hall conductivity and the pinning of fractional charges. The protocols proposed here may thus enable the observation of novel signatures of quantum Hall topological order in near-term quantum simulation platforms.  Moreover, our approach may be applied to other Floquet-engineered correlated systems~\cite{Kalinowski2023, Sun2023, Petiziol2024toriccode, Petiziol2024anyons} for the exploration of topological order in non-equilibrium driven-dissipative scenarios.

\begin{acknowledgments}
We thank Iacopo Carusotto for insightful discussions. This work has been supported by the Deutsche Forschungsgemeinschaft (DFG, German Research Foundation) via the Research Unit FOR 5688 (project No.~521530974).
F.~P.~and L.~S.~acknowledge funding from the Deutsche Forschungsgemeinschaft (DFG, German Research Foundation) through the Emmy Noether Programme – project number 555842149.
\end{acknowledgments}

\bibliography{main}

\onecolumngrid
\clearpage
\appendix

\begin{center}
   \hypertarget{endmatter}{\textbf{End Matter}}
\end{center}
\twocolumngrid
\paragraph*{Effective Lindblad equation.} In this section, we provide more details about how the model of \cref{eq:ELE} is derived. The system consists of an array of superconducting qubits, e.g., capacitively-coupled transmons, individually coupled to microwave resonators (cavities), as realized, for instance, in Ref.~\cite{Rosen2024}. The decay rate $\kappa_j$ of the resonators is assumed to be much stronger than any other dissipative rates, such as transmon dephasing and spontaneous decay, which are thus neglected, in first approximation, over the stabilization timescales considered here. The dynamics of the superconducting circuit is described by the Lindblad equation 
\begin{equation}
\dot{\hat \rho} = -\frac{i}{\hbar}[\hat H (t), \hat \rho] + \sum_j \kappa_j \mathcal{D}_{\hat c_j}(\hat \rho). 
\end{equation}
The Hamiltonian $\hat H(t)=\hat{H}_{\mathrm{s}}(t)+\hat{H}_{\rm c}(t) + \hat{H}_{\rm sc}(t)$ governs the composite system encompassing the qubit and resonator excitations, and their interaction, in the frame of the Floquet drive~\cite{petiziolCavityBasedReservoirEngineering2022}. It features the Hamiltonian 
\begin{equation}
\hat H_\text{s}(t) = - J \sum_{\langle l, l^\prime\rangle} e^{i[\theta_{l}(t)-\theta_{l^\prime}(t)]}\hat a_l^\dag \hat a_{l^\prime} + \text{H.c.}
\end{equation}
pertaining to the qubit excitations, which are described by hard-core bosonic ladder operators $\hat{a}_l$ and $\hat{a}_l^\dagger$. The $T$-periodic Floquet drive appears in the time-dependent Peierls phase $\theta_{l}(t)$, where $\hbar\dot\theta_l(t)$ corresponds to a periodic modulation of the qubits transition frequencies (on-site potentials in the Bose-Hubbard model). For a site $(m,n)$ of the square lattice, with linear index $l=Lm + n$, we choose 
\begin{equation}
\hbar\dot\theta_l(t)=m\hbar\omega + \lambda\sin\left(\omega t -\frac{\pi}{2}(m+n)\right),
\end{equation}
such that the Hamiltonian averaged over one period, $(1/T)\int_0^T \! dt \ \hat{H}_{\rm s}(t)$ with $T=2\pi/\omega$, matches the HHBH Hamiltonian of Eq.~\eqref{eq: HHBH Hamiltonian} in the high-frequency regime, $\hbar\omega\gg J$.
Excitations of the pumped resonators are created by operators $\hat{c}_j^\dag$ and are described by the Hamiltonian
\begin{equation}
\hat H_\text{c} = \sum_j \delta_j \hat c_j^\dag \hat c_j + \mathcal{E}_j^* e^{-i\omega_j t} \hat c_j + \mathcal{E}_j e^{i\omega_j t} \hat c_j^\dag,
\end{equation}
where $\delta_j$ is the qubit-cavity detuning for the cavity coupled to the qubit at site $j$. $\mathcal{E}_j$ and $\omega_j$ denote the pump amplitude and frequency, respectively. Qubits and resonators interact via a Jaynes-Cummings coupling, 
\begin{equation}
\hat H_\text{sc} = -\sum_j g_j e^{-i\theta_{j}(t)}\hat a_j\hat c_j^\dag + \text{H.c.},
\end{equation}
which is also time-dependent, via $\theta_j(t)$, in the frame of the drive, due to the modulation of the on-site potentials. 
The array-cavities coupling is assumed to be dispersive, namely, the cavities frequencies are far detuned from the qubit's single-excitation frequency, such that processes that exchange excitations between lattice and cavities are energetically frozen (in the absence of the Floquet drive). This guarantees that the total excitation number $\sum_l \hat{a}_l^\dag \hat{a}_l$ in the array is conserved, enabling quantum simulation at conserved particle number as desired. (Note that a small loss rate can be compensated easily by using post selection~\cite{petiziolCavityBasedReservoirEngineering2022}.) In this regime, the Jaynes-Cummings coupling translates into an effective density-density (cross-Kerr) coupling $\sim \hat{a}_j^\dag \hat{a}_j \hat{c}^\dag_j \hat{c}_j$, which can be derived in second-order perturbation theory~\cite{Blais2021} and lays the ground for the reservoir engineering approach adopted here~\cite{Hacohen-Gourgy2015}. However, this picture must be refined to account for the presence of the Floquet drive. Indeed, the latter can potentially restore (via accidental resonances) excitation tunneling between the array and the cavities, which may invalidate the dispersive approximation. Moreover, the drive alters the effective dispersive coupling. For an accurate description of bath engineering for the Floquet systems, these effects cannot be neglected. To account for them, we follow Ref.~\cite{petiziolCavityBasedReservoirEngineering2022} and derive an effective system-bath coupling by treating the high-frequency regime of the Floquet drive and the dispersive regime jointly in perturbation theory. We obtain the effective array-cavities Hamiltonian $\hat{H}_{\rm eff}= \hat{H}_{ \rm FCI} + \hat{H}_c + \hat{H}_{\rm sc}^{(\rm eff)}$, with 
\begin{align}
    \hat{H}_\text{sc}^\text{(eff)} = - \sum_j  \chi_j \left[ \hat n_j\hat c_j ^\dag \hat c_j + \frac{1}{2}\hat n_j - \frac{1}{2}\hat c_j^\dag \hat c_j\right],
\end{align}
where $\chi_j$ corresponds to the effective lattice-cavity coupling of the main text. In second-order perturbation theory, one also obtains additional dispersive-like shifts of the on-site potentials of $\hat{H}_{\rm FCI}$ and $\hat{H}_{c}$ on the order of $\chi_j$, which add disorder to the lattice: We assume that they are compensated for by an appropriate shift of the cavity-pump detuning $d_j = \delta_j - \omega_j$ and thus neglect them in $\hat{H}_{\rm eff}$. 
Finally, to arrive at the effective Lindblad equations stated in the main part (cf. \cref{eq: eff coupling} and above), we transform the cavity operators to a frame rotating at the pump frequency $\hbar\omega_j$ and displace them according to $\hat c_j \mapsto \hat c_j + \alpha_j $, where $\alpha_j$ is the amplitude of the coherent state the cavity would relax to if uncoupled from the lattice, 
\begin{equation}
\label{eq:alphaj}
\alpha_j=- \frac{\mathcal{E}_j}{d_j - \frac{i\hbar\kappa_j}{2}}
\end{equation}
As for the dispersive shifts, any weak on-site potential shifts induced by the cavity field displacements are assumed to be compensated for, to avoid disorder. 
\paragraph*{Pauli rate equation.}
 This section recapitulates the derivation of \cref{eq:PRE} from Eq.~\eqref{eq:ELE}, which is used to simulate the lattice dynamics after eliminating the cavity degrees of freedom, for system sizes at which solving Eq.~\eqref{eq:ELE} becomes computationally prohibitive. For dissipation engineering, the cavities are assumed to operate in the `bad-cavity' regime, where their relaxation rate $\kappa_j$ is much larger than the effective coupling $|\alpha_j \chi_j \bra{\varepsilon_\eta}\hat{n}_k\ket{\varepsilon_{\eta'}}|$ to the atoms, while both are much smaller than the relevant (quasi)energy difference $\varepsilon_\eta-\varepsilon_\eta'$ that the cavity is supposed to extract (or inject) [i.e., the level spacing that is on resonance with the cavity-pump detuning, $\varepsilon_\eta-\varepsilon_{\eta'}\approx d_j$], \begin{equation}
 \vert \varepsilon_{\eta} - \varepsilon_{\eta^\prime} \vert \gg \hbar \kappa_j \gg \vert \alpha_j\chi_j \bra{\varepsilon_\eta}\hat n_j\ket{\varepsilon_{\eta^\prime}}\vert.
 \end{equation}
 In this limit, the separation of timescales between the cavity and lattice relaxation allows one to treat the cavities as stationary in their steady state $\ket{\alpha_j}$ while the lattice evolves and, thus, as effective Markovian baths to the lattice. The stabilization dynamics of the eigenstate populations can then be cast into a classical Pauli rate equation between eigenstates $\ket{\varepsilon_\eta}$, with rates derived by following the Born-Markov-secular treatment of ultra-weak system-bath interactions~\cite{Breuer2002,Carmichael1993}. To illustrate this, consider the effective Hamiltonian $\hat{H}_{\rm eff}$ in the interaction picture and in the $\ket{\varepsilon_\eta}$ basis for the lattice, reading as
\begin{align} \label{eq:HeffEM}
    \hat H_\text{eff}(t) = - \sum_j  \chi_j \sum_{\eta, \eta^\prime} \bra{\varepsilon_\eta}\hat n_j\ket{\varepsilon_{\eta^\prime}}\ket{\varepsilon_\eta}\bra{\varepsilon_{\eta^\prime}} \hat O^{\eta\eta^\prime}_j(t),
\end{align}
with 
\begin{align}
    \hat O^{\eta\eta^\prime}_j(t) =& \hat c_j ^\dag \hat c_j e^{\frac{i}{\hbar}(\varepsilon_\eta-\varepsilon_{\eta^\prime}) t } + \alpha_j \hat c_j^\dag e^{\frac{i}{\hbar}(\varepsilon_\eta-\varepsilon_{\eta^\prime} +d_j) t}\nonumber \\ 
    &+ \alpha_j^* \hat c_j e^{\frac{i}{\hbar}(\varepsilon_\eta-\varepsilon_{\eta^\prime}  - d_j) t}.
\end{align}
Consider fixed $\eta$ and $\eta'$ and assume $\eta-\eta'>0$ and $d_j$ positive (red-detuned pump) for a given ($j$th) cavity. Setting the cavity-pump detuning $d_j$ to match the energy transition in the lattice, $d_j = \varepsilon_{\eta^\prime} - \varepsilon_\eta$, brings the photon creation term $\alpha_j \hat c_j^\dag$ on resonance, while the other two terms oscillate fast and average out. Effectively, these terms describe the fact that, in the Hamiltonian \eqref{eq:HeffEM}, photon absorption in the driven cavity is always accompanied by a transition $\ket{\varepsilon_{\eta^\prime}} \mapsto \ket{\varepsilon_\eta}$ in the lattice, such that the cavity pump and decay can be used to extract energy from the latter. This provides the basic processes for the engineered cooling pushing the system towards the ground state of $\hat{H}_{\rm FCI}$. Conversely, $\hat O^{\eta^\prime\eta}_j(t)$ contains resonant terms $\sim\hat c_j$ describing the transfer of energy from the cavity to the lattice (``heating''): to induce cooling, these processes are suppressed by the cavity's quantum noise spectrum, provided the cavity linewidth $\kappa_j$ is much narrower than $\varepsilon_\eta - \varepsilon_{\eta'}$. 
Following the Born-Markov-secular formalism~\cite{Breuer2002,Carmichael1993}, the transition between two lattice eigenstates $\ket{\varepsilon_{\eta^\prime}} \mapsto \ket{\varepsilon_\eta}$ occurs at a rate 
\begin{equation}
R_{\eta\eta^\prime} = \frac{2\pi}{\hbar}\sum_j \abs{\bra{\eta}\hat n_j \ket{\eta^\prime}}^2 \mathfrak{Re} W_j(\varepsilon_{\eta^\prime} - \varepsilon_\eta).
\end{equation}
Here, $W_j(\varepsilon) = \frac{1}{\pi\hbar}\int_0^\infty d\tau e^{-i\varepsilon/\hbar t} C(\tau)$ is the imaginary Laplace transform of the two-time correlator $C(\tau) = \langle \hat B^\dag(t+\tau) B(t)\rangle$ for the cavity coupling operator $\hat B_j = \chi_j\alpha_j \hat c_j^\dag$. The bath correlation function $C(\tau)$ of a pumped, leaky cavity can be evaluated using the quantum regression theorem~\cite{Carmichael1993} and reads 
\begin{equation}
C(\tau) = \chi_j^2\vert \alpha_j\vert^2 e^{-(id_j/\hbar + \kappa_j/2)\tau},
\end{equation}
where $\vert \alpha_j\vert^2$ is the photon number in the stationary state of the cavity [see Eq.~\eqref{eq:alphaj}]. Evaluating its imaginary Laplace transform yields the transition rates given in \cref{eq: pauli rates}. The Lorentzian shape of the transition rates reflect the fact that, in the bad-cavity regime, each cavity acts as narrowband filter to the surrounding environment. 
Note that, in the secular treatment leading to  \cref{eq: pauli rates}, the dynamics of the populations and coherences are decoupled from each other and the coherences eventually decay to zero in the steady state~\cite{Breuer2002}. Moreover, if the system departs from the bad-cavity limit, we expect that the transitions rates remain a good qualitative measure of the steady state~\cite{Murch2012}, as also seen in \cref{fig:collage_4x4}.

\paragraph*{Parameter ranges.} In this section, we discuss the parameter values used in numerical simulations. 
The complete list for all Figs.\ in the main text can be found in the Supplementary Material~\cite{Note1}.
To ensure that the HHBH Hamiltonian of Eq.~\eqref{eq: HHBH Hamiltonian} is a good approximation to the effective Floquet Hamiltonian in the lattice for high driving frequency $\omega$, we choose $\hbar\omega = 20 \, J$. The amplitude of the Floquet drive $\lambda$ is chosen such that the effective tunneling rate $J_{\rm eff}$ is uniform across the lattice, requiring $\lambda/\hbar\omega \approx 0.72/\sin({\phi/2})$ such that $J_\text{eff}/ J \approx 0.55$ for $\phi = \pi/2$~\cite{petiziolCavityBasedReservoirEngineering2022} (which is the case for all simulations except the results in \cref{fig:collage_4x4}c, where we scan the value of $\phi$ and thus adapt $\lambda$ accordingly). The remaining parameters are chosen to maximize (empirically) the ground state fidelity within the Floquet-dispersive regime $\abs{\delta_j + \mu\hbar\omega} \gg g_j \mathcal{J}_\mu \left(\frac{\lambda}{\hbar\omega}\right)$ and the bad-cavity limit $\hbar\kappa_j \gg \abs{\alpha_j\chi_j \bra{\eta}\hat n_j\ket{\eta^\prime}}$, while ensuring that unwanted Floquet multi-photon resonances are avoided. All values used are compatible with those implemented in superconducting-circuit experiments of the type considered here~\cite{Murch2012, Hacohen-Gourgy2015, Rosen2024, Wang2024}, assuming tunneling rates on the order of $J/2\pi\hbar \sim 20 - 50 \, \mathrm{MHz}$.

\cleardoublepage
 \let\oldaddcontentsline\addcontentsline
\renewcommand{\addcontentsline}[3]{}

\let\addcontentsline\oldaddcontentsline

\beginsupplement
\setcounter{page}{1}
\onecolumngrid
\begin{center}
{\bf \large Supplemental Material to} \\
\vspace{0.2cm}
{\bf \large Dissipation-assisted preparation of Floquet-Laughlin states in superconducting circuits} \\
\vspace{0.4cm}

Luis C. Steinfadt, André Eckardt, Francesco Petiziol \\

{\itshape
Institut für Physik und Astronomie, Technische Universität Berlin,
Hardenbergstr.\ 36, D-10623 Berlin, Germany}
\end{center}

\onecolumngrid  
\begin{center}\begin{table}[ht]
    \centering
    \caption{Parameters corresponding to the data presented in \cref{fig:collage_4x4}. Parameters are rounded to two significant digits if not specified otherwise. The first column indicates the corresponding figure/panel and the second column the sites to which the cavities are coupled with respect to the mapping $j = 4m + n$ for a site $(m,n)$ of the square lattice. The remaining columns show the cavity parameters for each of the used cavities. If a parameter is only given once, it is the same for all four cavities.}
    \begin{tabular}{|c|c|c|c|c|c|c|}
        \multicolumn{7}{c}{}\\[-1pt]
         \multicolumn{7}{c}{\cref{fig:collage_4x4}}\\
        \hline
         panel & site $j$ & $g_j/J$ & $\delta_j/\hbar\omega$ & $\mathcal{E}_j/J$ & $\kappa_j/J$ & $d_j/J$ \\
        \hline\multicolumn{7}{c}{}\\[-9pt]\hline
         a  & $1, 5, 15, 4$ & $0.8$ & $1.9, 1.85, 1.43, 1.9$ &  $0.4, 0.5, 1.0, 0.5$ & $0.04, 0.05, 0.1, 0.05$ & $0.13, 0.17, 0.4, 0.15$ \\
        \hline
        \multicolumn{7}{c}{}\\[-9pt]
        \hline
         b & $1, 0, 4, 5$ & $0.8$ & $1.9, 1.85, 1.57, 1.9$ & $0.4, 0.5, 0.7, 0.4$ & $0.04, 0.05, 0.07, 0.04$ &$ 0.13, 0.16, 0.26, 0.14$ \\
        \hline\multicolumn{7}{c}{}\\[-9pt]\hline
         c $\frac{\phi}{2\pi}< 0.275$ & \multicolumn{6}{c|}{cf. a} \\
        \hline
        $\frac{\phi}{2\pi}\ge0.275$ &  \multicolumn{4}{c|}{cf. a} &  $0.03, 0.05, 0.1, 0.05$ & $0.1, 0.17, 0.4, 0.15$ \\
        \hline\multicolumn{7}{c}{}\\[-1pt]
        \end{tabular}
\label{table}

\end{table}\end{center}

\begin{table}[ht]
\caption{Parameters corresponding to the data presented in \cref{fig:pinning}b. The first column indicates the corresponding value of the potential $V/J_\text{eff}$ where $J_\text{eff}$ is the Floquet-renormalized tunneling amplitude with $J_\text{eff} = 0.55\,J$. The linear site index is given with respect to the mapping $j = 8m + n$ for a site $(m,n)$ of the square lattice.}
\centering
\begin{tabular}{|c|c|c|}
\multicolumn{3}{c}{}\\[-1pt]
\multicolumn{3}{c}{\cref{fig:pinning}b (Part I: cavity sites and coupling) }\\
\hline
$V/J_{\text{eff}}$ & site $j$ & $g_j/J_{\text{eff}}$ \\
\hline
\hline
0 -- 0.375 & $9, 17, 10, 25, 11, 18, 1, 8$ & $0.37, 0.31, 0.31, 0.45, 0.45, 0.4, 0.6, 0.6$ \\
\hline
0.5 -- 1.875 & cf. $V/J_{\text{eff}}=0$ & $0.37, 0.31, 0.31, 0.45, 0.45, 0.4, 0.67, 0.67$ \\
\hline
2.0, 2.375, 2.5 & cf. $V/J_{\text{eff}}=0$ & $0.37, 0.31, 0.31, 0.45, 0.45, 0.45, 0.75, 0.75$ \\
\hline
2.125 & $9, 17, 10, 25, 11, 18, 1, 8, 26, 19$ & $0.4, 0.31, 0.31, 0.5, 0.5, 0.31, 0.83, 0.83, 0.35, 0.35$ \\
\hline
\end{tabular}
\begin{tabular}{|c|c|c|}
\multicolumn{3}{c}{}\\[-1pt]
\multicolumn{3}{c}{\cref{fig:pinning}b (Part II: reservoir detuning and cavity decay) }\\
\hline
$V/J_{\text{eff}}$ & $\delta_j/\hbar\omega$ & $\kappa_j/J_{\text{eff}}$ \\
\hline
\hline
$\neq 2.125$ & $1.83, 2.83, 2.83, 1.73, 1.73, 1.83, 1.83, 1.83$ & $0.024, 0.015, 0.015, 0.04, 0.04, 0.023, 0.023, 0.023$ \\
\hline
2.125 & $1.83, 2.83, 2.83, 1.73, 1.73, 1.83, 1.83, 1.83, 1.83, 1.83$ & $0.024, 0.015, 0.015, 0.04, 0.04, 0.016, 0.023, 0.023, 0.016, 0.016$ \\
\hline
\end{tabular}
\begin{tabular}{|c|c|c|}
\multicolumn{3}{c}{}\\[-1pt]
\multicolumn{3}{c}{\cref{fig:pinning}b (Part III: cavity pump amplitude and detuning) }\\
\hline
$V/J_{\text{eff}}$ & $\mathcal{E}_j/J_{\text{eff}}$ & $d_j/J_{\text{eff}}$ \\
\hline
\hline
$\neq 2.125$ & $0.4, 0.4, 0.4, 0.8, 0.8, 0.4, 0.4, 0.4$ & $0.0723, 0.05, 0.05, 0.135, 0.135, 0.0723, 0.0723, 0.0723$ \\
\hline
2.125 & $0.4, 0.4, 0.4, 0.8, 0.8, 0.4, 0.4, 0.4, 0.4, 0.4$ & $0.0723, 0.05, 0.05, 0.135, 0.135, 0.05, 0.0723, 0.0723, 0.05, 0.05$ \\
\hline
\end{tabular}
\end{table}

\begin{table}[ht]
\caption{Parameters corresponding to the data presented in \cref{fig:pinning}a. The first column indicates the corresponding figure/panel and the second column the sites to which the cavities are coupled with respect to the mapping $j = 6m + n$ for a site $(m,n)$ of the square lattice. The remaining columns show the cavity parameters for each of the used cavities. If a parameter is only given once, it is the same for all cavities. \text{*}For $V/J = 0.6$ and $0.65$, all parameters are identical except for the final element in $g_j/J$, which is $0.4$ for $V/J=0.6$ and $0.45$ for $V/J=0.65$. They have been grouped here showing $0.4$ only. }
\centering

\begin{tabular}{|c|c|c|c|}
\multicolumn{4}{c}{}\\[-.3pt]
\multicolumn{4}{c}{\cref{fig:pinning}a (Part I: cavity sites, decay, pump detuning) }\\
\hline
$V/J$ & site $j$ & $\kappa_j/J$ & $d_j/J$ \\
\hline
\hline
0 -- 0.05 & $7, 13, 8, 1, 6, 2, 12$ & $0.021, 0.03, 0.03, 0.02, 0.06, 0.021, 0.021$ & $0.062, 0.09, 0.09, 0.062, 0.18, 0.062, 0.062$ \\
\hline
0.1 -- 0.15 & cf. $V/J=0$ & $0.018, 0.03, 0.03, 0.018, 0.06, 0.021, 0.021$ & $0.055, 0.09, 0.09, 0.055, 0.18, 0.062, 0.062$ \\
\hline
0.2 & $7, 13, 8, 1, 6, 2, 12, 14$ & $0.018, 0.03, 0.03, 0.018, 0.06, 0.021, 0.021, 0.023$ & $0.055, 0.09, 0.09, 0.055, 0.18, 0.062, 0.062, 0.07$ \\
\hline
0.25 & $7, 13, 2, 8, 6, 1, 12, 14$ &  \multicolumn{2}{c|}{cf. $V/J=0.2$} \\
\hline
0.3 & cf. $V/J=0.25$ & $0.01, 0.03, 0.03, 0.01, 0.06, 0.021, 0.021, 0.016$ & $0.03, 0.09, 0.09, 0.03, 0.18, 0.062, 0.062, 0.05$ \\
\hline
0.5 & $7, 13, 8, 1, 6, 14, 2$ & $0.011, 0.02, 0.02, 0.04, 0.04, 0.03, 0.033$ & $0.035, 0.06, 0.06, 0.12, 0.12, 0.09, 0.11$ \\
\hline
0.55 & $7, 13, 8, 1, 6, 14$ & $0.011, 0.02, 0.02, 0.04, 0.04, 0.03$ & $0.035, 0.06, 0.06, 0.12, 0.12, 0.091$ \\
\hline
0.6 -- 0.65 & $7, 14, 8, 1, 6, 13$ & $0.014, 0.027, 0.027, 0.04, 0.04, 0.035$ & $0.042, 0.085, 0.085, 0.12, 0.12, 0.105$ \\
\hline
0.70 & cf. $V/J=0.6$ & $0.0143, 0.028, 0.028, 0.04, 0.04, 0.046$ & $0.044, 0.085, 0.085, 0.13, 0.13, 0.14$ \\
\hline
0.75 & cf. $V/J=0.6$ & $0.0143, 0.027, 0.027, 0.05, 0.05, 0.053$ & $0.044, 0.084, 0.084, 0.15, 0.15, 0.16$ \\
\hline
0.8 & cf. $V/J=0.6$ & $0.0143, 0.027, 0.027, 0.05, 0.05, 0.06$ & $0.044, 0.084, 0.084, 0.15, 0.15, 0.18$ \\
\hline
0.85 & cf. $V/J=0.6$ & cf. $V/J=0.8$ & $0.044, 0.084, 0.084, 0.18, 0.18, 0.23$ \\
\hline
0.9 & $7, 14, 2, 1, 6, 13$ & $0.013, 0.027, 0.027, 0.05, 0.05, 0.06$ & $0.04, 0.08, 0.08, 0.20, 0.20, 0.25$ \\
\hline
\end{tabular}
\begin{tabular}{|c|c|c|c|}
\multicolumn{4}{c}{}\\[-1pt]
\multicolumn{4}{c}{\cref{fig:pinning}a (Part II: cavity coupling, reservoir detuning, cavity pump amplitude) }\\
\hline
$V/J$ & $g_j/J$ & $\delta_j/\hbar\omega$ & $\mathcal{E}_j/J$ \\
\hline
\hline
0 & $0.35, 0.4, 0.4, 0.35, 0.8, 0.45, 0.45$ & $1.8$ & $0.36, 0.5, 0.5, 0.36, 0.73, 0.33, 0.33$ \\
\hline
0.05 & $0.35, 0.4, 0.4, 0.38, 0.8, 0.4, 0.4$ & cf. $V/J=0$ & cf. $V/J=0$ \\
\hline
0.1 & $0.3, 0.42, 0.42, 0.34, 0.8, 0.43, 0.43$ & cf. $V/J=0$ & cf. $V/J=0$ \\
\hline
0.15 & $0.3, 0.42, 0.42, 0.34, 0.8, 0.46, 0.46$ & cf. $V/J=0$ & cf. $V/J=0$ \\
\hline
0.2 & $0.3, 0.4, 0.4, 0.34, 0.8, 0.42, 0.42, 0.3$ & cf. $V/J=0$ & $0.36, 0.5, 0.5, 0.36, 0.73, 0.33, 0.33, 0.33$ \\
\hline
0.25 & $0.3, 0.4, 0.5, 0.3, 0.8, 0.42, 0.42, 0.34$ & cf. $V/J=0$ & cf. $V/J=0.2$ \\
\hline
0.3 & $0.2, 0.4, 0.5, 0.2, 0.8, 0.42, 0.42, 0.3$ & cf. $V/J=0$ & $0.25, 0.5, 0.5, 0.25, 0.73, 0.33, 0.33, 0.2$ \\
\hline
0.5 & $0.23, 0.4, 0.4, 0.5, 0.5, 0.3, 0.5$ & $1.8, 1.8, 1.8, 1.61, 1.61, 1.8, 1.8$ & $0.19, 0.23, 0.23, 0.6, 0.6, 0.5, 0.6$ \\
\hline
0.55 & $0.23, 0.43, 0.43, 0.5, 0.5, 0.3$ & $1.8, 1.8, 1.8, 1.61, 1.61, 1.8$ & $0.19, 0.23, 0.23, 0.6, 0.6, 0.5$ \\
\hline
0.6 -- 0.65 & $0.27, 0.3, 0.4, 0.53, 0.53, 0.4$\textsuperscript{*} & $1.8, 1.8, 1.8, 1.61, 1.61, 1.69$ & $0.2, 0.4, 0.51, 0.6, 0.6, 0.55$ \\
\hline
0.70 & $0.27, 0.27, 0.4, 0.53, 0.53, 0.5$ & $1.8, 1.8, 1.8, 1.67, 1.67, 1.67$ & $0.23, 0.51, 0.51, 0.7, 0.7, 0.8$ \\
\hline
0.75 & $0.25, 0.25, 0.4, 0.53, 0.53, 0.5$ & cf. $V/J=0.70$ & $0.27, 0.51, 0.51, 0.9, 0.9, 1$ \\
\hline
0.8 & $0.3, 0.25, 0.45, 0.56, 0.56, 0.53$ & cf. $V/J=0.70$ & $0.2, 0.51, 0.4, 0.9, 0.9, 1.2$ \\
\hline
0.85 & $0.3, 0.25, 0.45, 0.67, 0.67, 0.6$ & cf. $V/J=0$ & cf. $V/J=0.8$ \\
\hline
0.9 & $0.26, 0.23, 0.47, 0.69, 0.69, 0.63$ & cf. $V/J=0.85$ & cf. $V/J=0.8$ \\
\hline
\end{tabular}
\end{table}
\end{document}